\documentclass[superscriptaddress,preprint,amsmath]{revtex4}

\usepackage{graphicx}

\begin{document}


\title{Controllable optical phase shift over one radian from a single isolated atom}

\author{A. Jechow}
\author{B. G. Norton}
\author{S. H{\"a}ndel}
\author{V. Bl{\=u}ms}
\author{E. W. Streed}
\author{D. Kielpinski}
\email[email:]{ d.kielpinski@griffith.edu.au }
\affiliation{Centre for Quantum Dynamics, Griffith University, Brisbane 4111, QLD, Australia.}

\maketitle


\textbf{Fundamental optics such as lenses and prisms work by applying phase shifts to incoming light via the refractive index. In these macroscopic devices, many particles each contribute a miniscule phase shift, working together to impose a total phase shift of many radians. In principle, even a single isolated particle can apply a radian-level phase shift, but observing this phenomenon has proven challenging. We have used a single trapped atomic ion to induce and measure a large optical phase shift of $1.3 \pm 0.1$ radians in light scattered by the atom. Spatial interferometry between the scattered light and unscattered illumination light enables us to isolate the phase shift in the scattered component. The phase shift achieves the maximum value allowed by atomic theory over the accessible range of laser frequencies, validating the microscopic model that underpins the macroscopic phenomenon of the refractive index. Single-atom phase shifts of this magnitude open up new quantum information protocols, including long-range quantum phase-shift-keying cryptography \cite{Inoue-Yamamoto-quantum-DPSK-crypto-theory, Takesue-Yamamoto-lossy-phase-shift-QKD} and quantum nondemolition measurement \cite{Giovannetti-Maccone-quantum-enhanced-measurement-rev, Volz-Reichel-single-atom-nondemolition-measurement}.} \\


Optical phase shifts are commonly observed from all materials, and generally originate from the delayed response of electrons to an applied light field. In principle, large phase shifts persist down to the single-atom level. As the frequency of light is tuned through atomic resonance, semiclassical theory predicts that the scattered light experiences a phase advance of $0$ for far red detuning, through $\pi/2$ on resonance, to $\pi$ for far blue detuning. Small phase shifts have recently been observed from single atoms \cite{Aljunid-Kurtsiefer-single-atom-phase-shift} and molecules \cite{Pototschnig-Sandoghdar-phase-contrast-single-molecule}. Similar small phase shifts, arising from the shift of atomic energy levels, have also been observed by interferometric measurements of fluorescence from single trapped ions \cite{Wilson-Blatt-single-mirror-vacuum-shift}. The interferometric techniques used in those experiments detected only the on-axis interference between the scattered field and the illumination field, so the properties of the scattered field could not be studied in isolation. Since the illumination field is always much stronger than the scattered field in these configurations, the accessible phase shift was limited at the 100 mrad level by the scattering amplitude, restricting applications in quantum information processing and nanophotonics. In another recent approach, by confining an atom in a high-finesse optical cavity, the phase shift effect was magnified by the many passages of the illumination light through the atom \cite{Kerckhoff-Mabuchi-single-atom-cavity-PSK}. Here we demonstrate control of a radian-level phase shift of scattered light for an isolated atomic ion in free space, as promised by theory.

\begin{figure}[htbp]
\begin{center}
\includegraphics[width=89mm,bb=0 0 289 208]{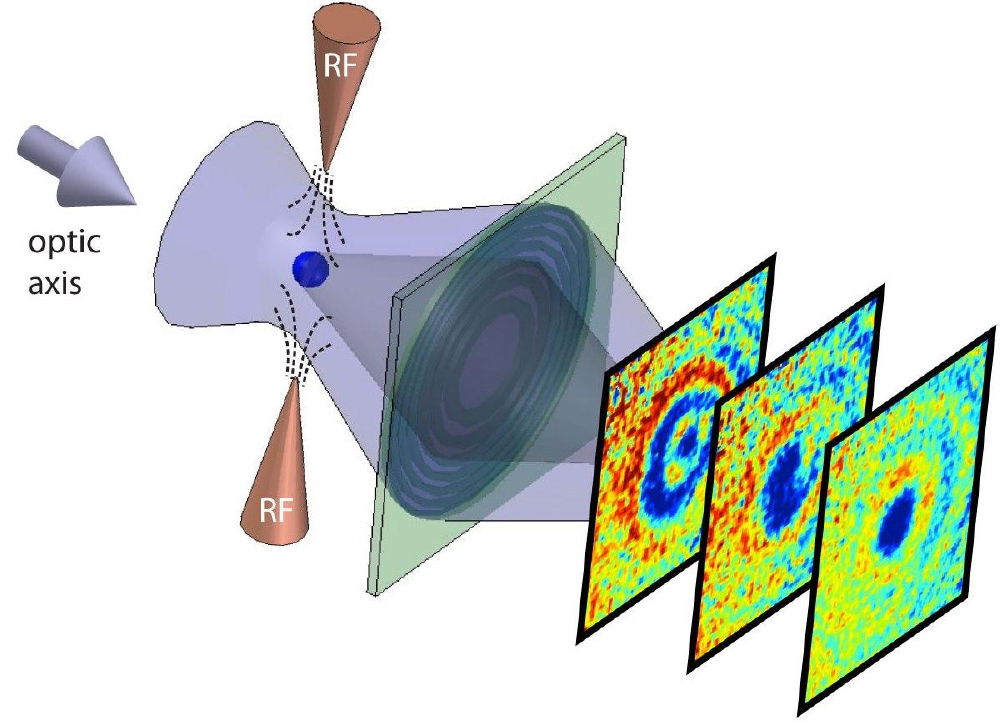}
\caption{Configuration of experimental apparatus. A laser cooled $^{174}$Yb$^+$ ion (blue dot) is confined in a radio-frequency electric quadrupole trap generated by two tungsten needles. Resonant laser light at $\lambda =369.5$ nm is incident along the optical axis of the imaging system and focused to a spot of $5 \:\mu$m FWHM at the ion position. The transmitted light, consisting of a superposition of the scattering and driving fields, is imaged with a large aperture phase Fresnel lens \cite{Streed-Kielpinski-Fresnel-ion-imaging, Jechow-Kielpinski-wavelength-scale-ion-imaging} onto a cooled CCD camera at $585\times$ magnification. We acquire images of the transmitted light at several camera viewing planes. A secondary cooling beam (not shown) is incident orthogonal to the needle axis and the optical axis.}
\label{exptschem}
\end{center}
\end{figure}

Our experimental apparatus is similar to that used in our recent work \cite{Jechow-Kielpinski-wavelength-scale-ion-imaging, Streed-Kielpinski-single-atom-absorption-imaging}. A schematic of the apparatus is shown in Figure \ref{exptschem}. A single $^{174}$Yb$^+$ ion is trapped in ultra-high vacuum using a double-needle radio-frequency (RF) quadrupole Paul trap operating at a drive frequency of 40 MHz. Laser light at 369.5 nm, near resonance with a strong transition of the ion, is weakly focused onto the ion to provide an illumination field with power of $\sim 5$ nW and spot diameter of $5 \: \mu$m (full-width at half-maximum). The illumination beam is linearly polarised to eliminate optical pumping effects and its power is actively stabilised to minimise intensity fluctuations between reference and signal images. The light transmitted past the atom is reimaged onto a cooled CCD camera with a magnification of $585$. To provide additional laser cooling, an auxiliary 369.5 nm laser beam, detuned $-200$ MHz from atomic resonance, is applied perpendicular to the optical axis. This additional cooling enables us to tune the illumination field somewhat blue of resonance while maintaining reasonable image contrast. The laser cooling ensures that the amplitude of the ion's motion is always much smaller than the imaging resolution. All data presented here have been obtained from a single continuously trapped ion over a period of a few hours.

Our data consist of background-subtracted, normalised images of the light transmitted past a single trapped ion, which amount to spatial interferograms of the light field scattered by the ion. These images are obtained by subtracting signal images, for which ion absorption was present, from reference images of the illuminating beam. Each pixel of the subtracted image is then normalised to its value in the reference image. To acquire the reference images, we optically pump the ion into the metastable $D_{3/2}$ atomic state, which scatters only a negligible amount of the 369.5 nm light. In the observation plane (i.e., the plane imaged onto the camera), the intensity of the scattered field $U_\mathrm{sc}(x,y)$ is everywhere much smaller than the intensity of the illumination field $U_0(x,y)$. Within the image area, the illumination field intensity is uniform and is independent of the observation plane position, so the phase and amplitude of the illumination field is uniform. The background-subtracted, normalised image signal, denoted $S(x,y)$, is therefore given by
\begin{equation}
S(x,y) \propto |U_0(x,y) + U_\mathrm{sc}(x,y)|^2 - |U_0(x,y)|^2 \approx 2 \:\mbox{Re} \left[ U_\mathrm{sc}(x,y) \: U_0(x,y) \right] \propto \mbox{Re} \left[ U_\mathrm{sc}(x,y) \right] \label{homodyne}
\end{equation}
The image signal $S(x,y)$ is seen to be a spatial interferogram of the scattered field, with the illumination field serving as the reference wave.

The high information content of the spatial interferogram $S(x,y)$ enables us to isolate the parameters of the scattered field $U(x,y)$, in particular the phase shift. The total transmitted field amplitude $U_0(x,y) + U_\mathrm{sc}(x,y)$ exhibits only small phase shifts relative to the illumination field in our data. Nevertheless, in order to match the spatial dependence of $S(x,y)$ to well-understood models of wave optics, we are constrained to assign large phase shifts to the scattered component of the total field. The imaging technique therefore accesses the scattered field alone, in contrast to results from previous spatially unresolved measurements \cite{Aljunid-Kurtsiefer-single-atom-phase-shift, Pototschnig-Sandoghdar-phase-contrast-single-molecule}. Since those previous measurements only probed the total transmitted power, the large-amplitude illumination field overwhelmed the small-amplitude scattered wave contribution. Hence the phase shift observed in those measurements was always reduced by a factor of $U_\mathrm{sc}/U_0 < 0.1$ relative to the results presented here, accounting for the previous observations of $\lesssim 100$ mrad phase shifts.

\begin{figure}[htbp]
\begin{center}
\includegraphics[width=89mm,bb=0 0 623 376]{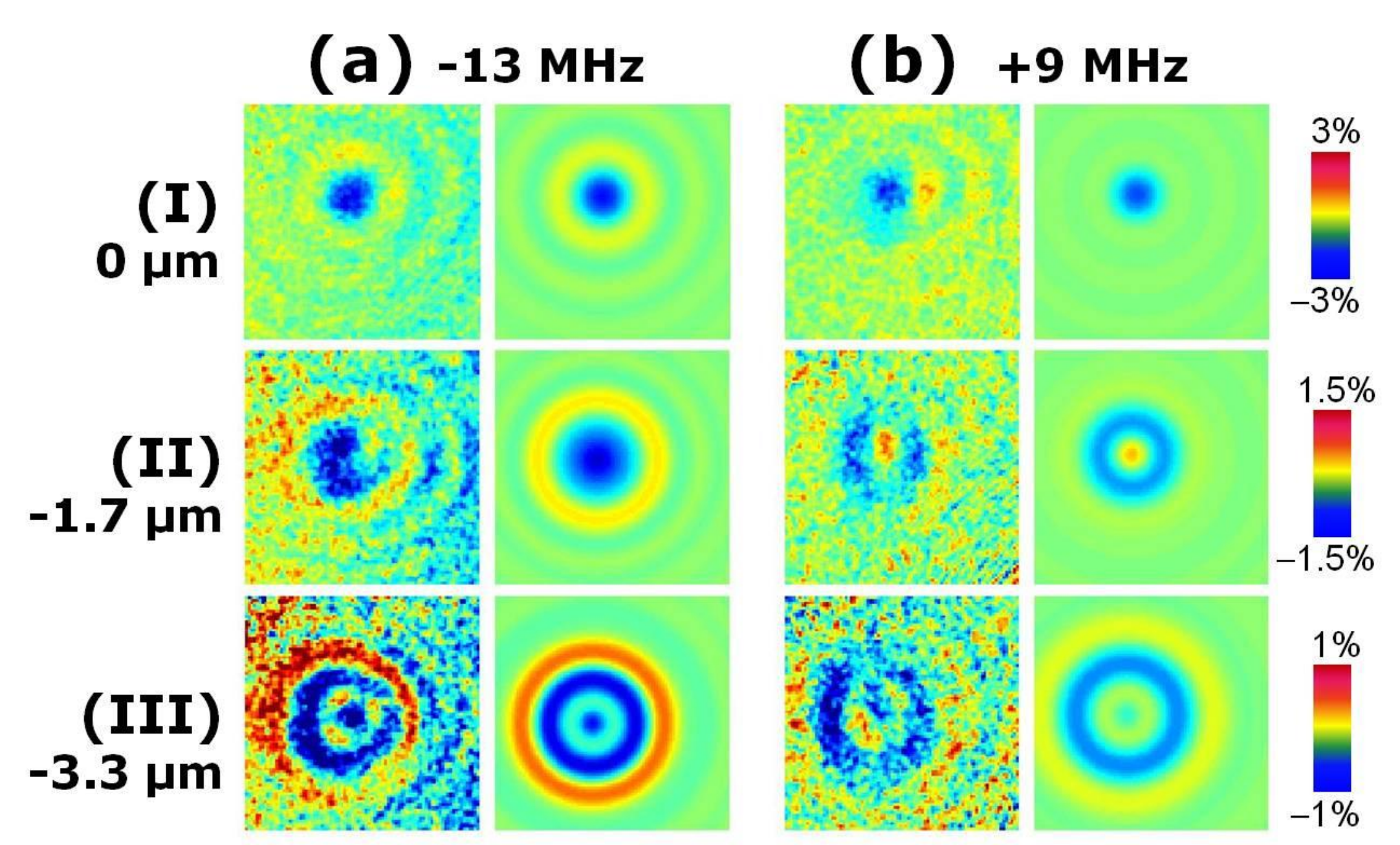}
\caption{Spatial interferograms of the scattered wave. The theoretical prediction for each interferogram is shown to the right of the data. The image resolution is 370 nm, approximately equal to the illumination wavelength of 369.5 nm, and each image is $3.4 \:\mu\mbox{m}$ on a side. Each row of images corresponds to a fixed observation plane position. In terms of the object space coordinates, row I sits at the nominal plane of the ion and row II (resp. III) at 1.7 $\mu$m (resp. 3.3 $\mu$m) upstream of the ion. The colorbars at the right of the figure indicate the fractional change in transmission for the images in each row. The images are smoothed with a Gaussian filter of 40 nm width for ease of viewing, but only raw data is used for comparison with theory. (a) Data (left) and theory (right) at $-13$ MHz detuning. (b) The same, but for $+9$ MHz detuning.}
\label{imagedata}
\end{center}
\end{figure}

Figure \ref{imagedata} shows a series of single-ion interferograms at different observation planes and laser detunings. When the imaging system is focused at the plane of the ion (row I in Fig. \ref{imagedata}), the scattered light always interferes destructively with the transmitted light, giving rise to an absorption image of the ion. The contrast of the absorption image, as well as the detuning dependence of the contrast, accords with the semiclassical theory of the atom-light interaction and with our previous measurements \cite{Streed-Kielpinski-single-atom-absorption-imaging}. For defocused imaging (rows II and III in Fig. \ref{imagedata}), the interference of the approximately spherical scattered wave with the planar illumination wave gives rise to a ``bullseye'' pattern. The extent of the bullseye grows with increasing defocusing as the scattered wave spreads transversely. The detuning-dependent phase shift of the scattered wave induces an alteration of the interference pattern, which is immediately evident from the on-axis intensity in row II and more subtly affects the spacing of the interference rings in row III.

We determine the phase of the scattered wave at each detuning from a series of interferogram images similar to those shown in Figure \ref{imagedata}. For each image series, the detuning is fixed and the observation plane is shifted to several positions along the optic axis. Each image series is fitted to a simulation of the wave propagation through our imaging system, and the scattered wave parameters are extracted from the fit. Typical fit images are shown in Figure \ref{imagedata}. When used to simulate fluorescence images, this model leads to good agreement with previous experimental data \cite{Jechow-Kielpinski-wavelength-scale-ion-imaging}. Further details of the model can be found in the Methods Summary.

Despite the presence of the auxiliary red-detuned cooling beam, we still incur reduced image contrast when the illumination beam is blue-detuned, but the reduced contrast does not affect our ability to measure the phase shift. The auxiliary cooling beam intensity is kept relatively low to avoid excessive saturation of the atomic transition, so the ion temperature still rises above the Doppler limit for blue detuning. The resulting thermal motion reduces the effective imaging resolution at blue detuning \cite{Norton-Kielpinski-ion-spatial-thermometry}. Along with residual saturation by the auxiliary beam, the lower resolution accounts for the reduction in contrast seen in Fig. \ref{exptschem}(b). The cooling beam also imparts an AC Stark shift to the atomic levels, which is calculated to be negligible for our experiment.

\begin{figure}[htbp]
\begin{center}
\includegraphics[width=85mm,bb=0 0 363 431]{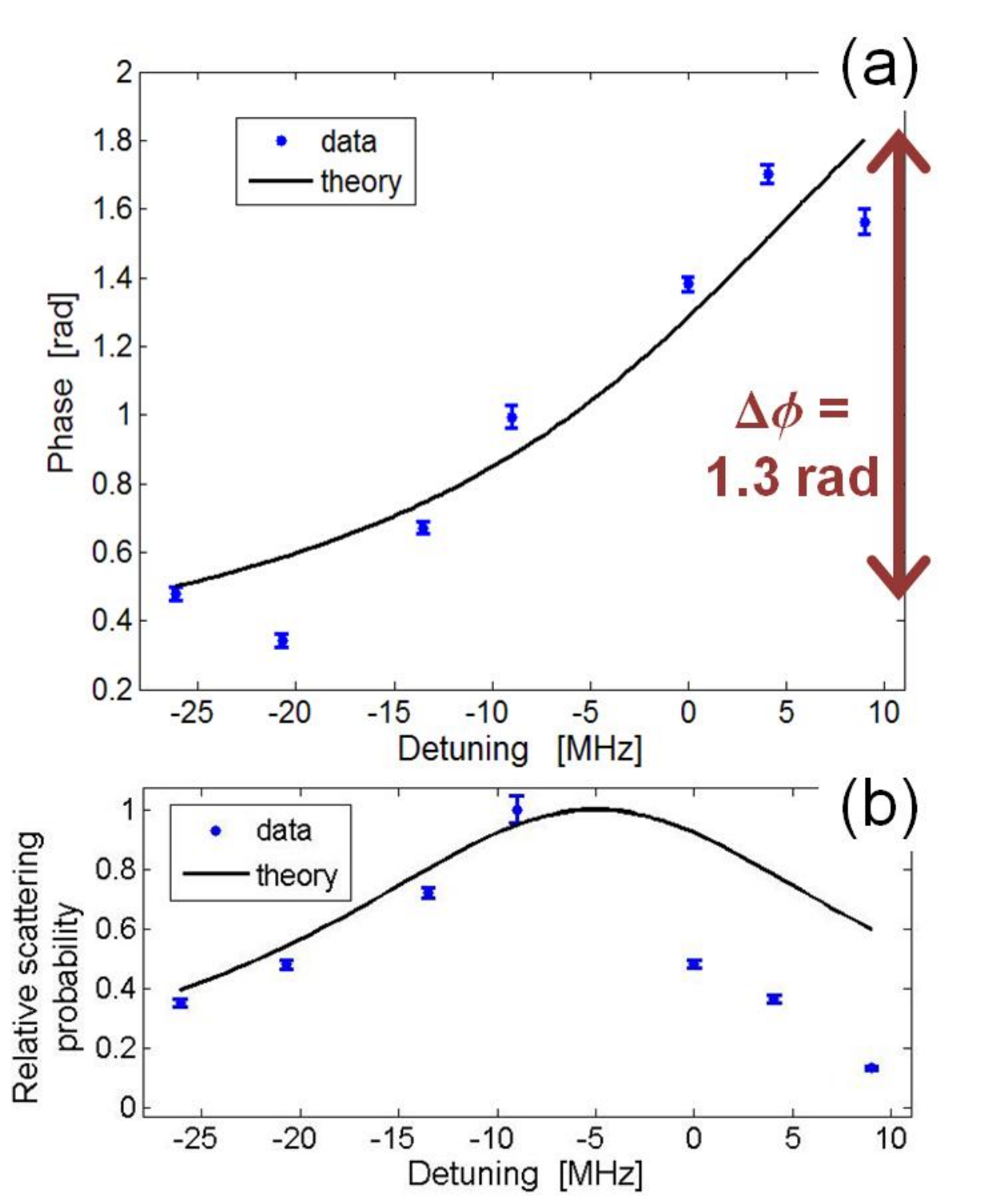}
\caption{Phase shift and normalised scattering probability of the scattered wave. a) Phase of the scattered wave as a function of laser detuning. Each data point is obtained by fitting a series of spatial interferograms to the model. The data shows a total phase shift of $1.3 \pm 0.1$ radians. The uncertainty in determining spherical aberration imparts a common systematic error of $\pm 0.1$ rad to all data points.  This common-mode error does not affect the measurement of the phase shift. The data is well fit by semiclassical theory with the linewidth $\Gamma = 34 \pm 8$ MHz and the fitted resonance position shifted $5 \pm 2$ MHz blue of the nominal resonance frequency. b) Normalised scattering probability as a function of detuning. All values are normalised to the maximum scattering probability observed in the data. The theory curve is predicted from the fit to the phase shift. On resonance and at blue detuning, the scattering probability is lower than expected from the theory. The mechanical effects of the laser light are seen to be significant, including the broadening and red-shift of the phase-shift data relative to the ideal case.}
\label{phaseshift}
\end{center}
\end{figure}

The retrieved phase of the scattered wave is shown as a function of detuning in Figure \ref{phaseshift}. Since the intensity of the illumination is kept well below the saturation intensity of the atomic transition ($600 \: \mbox{W}\:\mbox{cm}^{-2}$), the semiclassical theory of atomic scattering in a weak laser field should apply. The phase $\phi$ then depends only on the laser detuning $\Delta$ as follows:
\begin{equation}
\phi(\Delta) = \tan^{-1} (\Delta/\Gamma) + \pi/2
\end{equation}
where $\Gamma$ is the atomic frequency linewidth (full-width at half maximum). Ion heating to the blue side of atomic resonance is expected to increase the measured value of $\Gamma$ above the nominal 20 MHz exhibited by an ideal $^{174}$Yb$^{+}$ ion at rest. Similarly, our fluorescence measurement of the resonance frequency is expected to underestimate the actual frequency, since ion heating lowers the fluorescence rate rapidly to the blue of resonance. We fit the phase measurements to the semiclassical model and find $\Gamma = 34 \pm 8$ MHz with the fitted resonance position shifted $5 \pm 2$ MHz blue of the nominal resonance frequency. The close agreement with atomic theory evidences the near-ideal character of our system.

By showing that a single atom can impose radian-level phase shifts, our results bridge the gap between microscopic light scattering and macroscopic optical effects. Our results directly verify long-held theories about the microscopic origin of the refractive index. By adding atoms one at a time to our system, we can study the transition from microscopic quantum-optical phase shifting to macroscopic refractive optics. Our results also enable new protocols in quantum communication and measurement, since one may also manipulate the internal quantum states of the atom. For instance, in quantum phase-shift-keying (QPSK) cryptography, quantum information is encoded in the phase shift of a single photon \cite{Inoue-Yamamoto-quantum-DPSK-crypto-theory}, making it extremely resistant to decoherence and channel loss \cite{Takesue-Yamamoto-lossy-phase-shift-QKD}. The detuning dependence of the single-atom phase shift can be exploited to entangle the atomic state with QPSK photonic states, e.g., by using an atom with a Zeeman-split spin-1/2 ground state and driving atomic scattering with a laser tuned between the Zeeman-split transition components. The photonic states produced by atoms at remote nodes can then be used for long-distance QPSK cryptography by entanglement swapping, according to standard quantum repeater protocols \cite{Duan-Zoller-linear-optics-atom-QC}. Conditional phase shifting of multiple photons can also be used to read out the state of the atom in a minimally destructive way. Since all photons undergo the same conditional phase shift, the photonic state is entangled, permitting Heisenberg-limited estimation of the optical phase \cite{Giovannetti-Maccone-quantum-enhanced-measurement-rev} and therefore of the atomic state. In combination with the stable homodyne readout provided by our imaging technique, the measurement back-action can approach the limits achieved with much more challenging techniques based on optical resonators \cite{Volz-Reichel-single-atom-nondemolition-measurement}.

\vspace{1cm}
\textbf{Methods Summary}\\
\begin{small}

\textit{Details of fitting model.}

Our fitting model takes the scattered wave to be a scalar spherical wave. This assumption is well justified our numerical aperture (NA) of 0.64, since our geometry restricts the scattered polarisation to $\sigma^\pm$ along the imaging axis \cite{Streed-Kielpinski-ion-Fresnel-proposal}. The illumination field is modeled as a low-NA Gaussian beam with standard Gaussian beam propagation theory.

The spherical scattered wave is propagated nonparaxially up to the Fresnel lens and then propagated through the refocusing lens to the image plane using the Fresnel approximation. We model the Fresnel lens as a near-perfect lens with a small amount of spherical aberration and a super-Gaussian pupil function. The Fresnel lens is modeled as a thin complex transmittance. The transmittance phase removes nearly all of the off-axis spatial phase variation of the scattered spherical wave, leaving only the spherical aberration phase function $\Phi(\rho) = A \rho^4$, where $\rho$ is the distance from the optic axis in the plane of the Fresnel lens and $A$ quantifies the magnitude of spherical aberration. The transmittance amplitude is taken to be a super-Gaussian function $p_S \propto e^{-\rho^4/\rho_0^4}$, defining the diffraction-limited resolution of the imaging system through the pupil parameter $\rho_0$. The scattered wave is then reimaged onto the camera image plane by a weak lens: this process is modeled using the Fresnel diffraction integral, a valid approximation for this low-NA portion of the optical system. All told, the model predicts that the scattered field in the image plane $(x,y)$ is given by
\begin{align}
U_\mathrm{sc}(\xi) &= - U_0 a_\mathrm{sc} e^{i \phi_\mathrm{sc}} \frac{i e^{-i k (f_F + f_R)}}{\lambda f_R} \tilde{u}_\mathrm{sc} \left(\frac{\xi}{\lambda f_R} \right) \label{uscim} \\
\tilde{u}_\mathrm{sc} &\equiv {\cal F} \left[ \exp[i \pi \zeta \rho^2/(f_F^2 \lambda)] \frac{p_S(\rho)}{\sqrt{f_F^2 + \rho^2}} \right]
\end{align}
for small deviations of the viewing plane $\zeta$ along the optic axis, measured relative to the in-focus image plane. Here $\xi = \sqrt{x^2+y^2}$ is the transverse distance from the optic axis in the image-plane coordinates, the drive field amplitude is $U_0$, the atomic scattering amplitude and phase are $a_\mathrm{sc}$ and $\phi_\mathrm{sc}$, the Fresnel lens focal length is $f_F$, the reimaging lens focal length is $f_R$, and ${\cal F}[g]$ denotes the two-dimensional Fourier transform of a test function $g(x,y)$.

\vspace{1cm}
\noindent \textbf{Acknowledgements}
\noindent Supported by the Australian Research Council under FT110100513 (DK, Future Fellowship) and DP0877936 (EWS, Australian Postdoctoral Fellowship). AJ was supported by a Griffith University Postdoctoral Fellowship. The phase Fresnel lens was fabricated by M. Ferstl at the Heinrich-Hertz-Institut of the Fraunhofer-Institut f\"{u}r Nachrichtentechnik in Germany.

\vspace{1cm}
\noindent \textbf{Competing Interests}
The authors declare that they have no competing financial interests.

\vspace{1cm}
\noindent \textbf{Author contributions}
\noindent EWS and DK designed the experiment. AJ, BGN, VB, EWS and DK constructed the apparatus. AJ and BGN conducted the experiment and collected the data with the help of SH. DK and EWS analysed the data. The manuscript was prepared by AJ and DK with contributions from BGN, SH, VB, and EWS.

\vspace{1cm}
\noindent \textbf{Additional information}
\noindent Reprints and permissions information is available online at http://npg.nature.com/reprintsandpermissions. Correspondence and requests for materials should be addressed to DK~(email: d.kielpinski@griffith.edu.au).

\end{small}

\end{document}